\documentclass[twocolumn,preprintnumbers,amsmath,amssymb]{revtex4}

\usepackage{color}
\usepackage{relsize}
\usepackage{graphicx}
\usepackage{dcolumn}
\usepackage{bm}
\usepackage{wrapfig}
\usepackage{graphicx}

\newcommand{\be}{\begin{equation}}
\newcommand{\ee}{\end{equation}}
\newcommand{\ba}{\begin{array}}
\newcommand{\ea}{\end{array}}

\begin{document}


\title{An energetic model for macromolecules unraveling}

\author{D. De Tommasi$^1$, N. Millardi$^1$, G. Puglisi$^1$ and G. Saccomandi$^2$}
\affiliation{
$^{1}$Dipartimento di Scienze dell'Ingegneria Civile e dell'Architettura (ICAR), Politecnico
di Bari, Italy}

\affiliation{
$^{2}$Dipartimento di Ingegneria Industriale, Universit\`a degli Studi di Perugia}

\begin{abstract}
\noindent We propose a simple approach, based on the minimization of the total (entropic plus unfolding) energy of a two-state system, describing the stretch-induced unfolding of  macromolecules (proteins, silks, nanopolymers, DNA/RNA). The model is fully analytical and enlightens the role of the different energetic components regulating the unfolding evolution. As an explicit example of application we compare the analytical results with the titin Atomic Force Microscopy experiments showing the ability of the model to quantitatively reproduce the mechanical behavior of macromolecules unfolding.

\vspace{ 1 cm}


\end{abstract}

\maketitle

\section{Introduction}

The last decade has shown a significant theoretical and experimental effort  in the analysis on the thermo-mechanical behavior of macromolecular materials such as  muscle tissues, spider silks \cite{DeTommasi-Puglisi-Saccomandi2} \cite{Oroudjev}, polymers and biopolymers \cite{Harrington}, polysoaps \cite{Borisov1} \cite{Borisov2} and silks in general \cite{Hennady}.  A common property (see {\it e.g.} \cite{Harrington}, \cite{KM}, \cite{Miserez}, \cite{DeTommasi-Puglisi-Saccomandi2}, and \cite{Termonia}) is that their macroscopic history dependent, dissipative properties descend by complex ``semicrystalline''microstructures, constituted by flexible (polymeric or protein) macromolecules reinforced by strong and stiff crystals {\it e.g.} in form of fillers or $\beta$-sheets (see for example \cite{BY} \cite{Bueler}).

At small strain the stiffness of these material is mainly regulated by the hard fraction (secondary structure). Under stretching a reversible hard-soft transition (e.g. $\beta$-sheets unfolding in protein or  crosslinks breakage in polymers) is observed. This transition has the two important effects of dissipating energy , due to the transition itself, and of a variation of the microstructure, leading to variable contour lengths of the chains. At larger stretches, the behavior is regulated by the entropic hardening of the macromolecules (primary structure).
As a result, macroscopically these materials shows a hysteretic behavior, reminiscent of the pseudoelastic behaviour characterizing materials such as Shape Memory Alloys (see \cite{ Puglisi-Truskinovsky1} and references).

The deduction of predictive models that connect the mesoscale properties with the macroscopic material response are crucial not only to describe the behavior of such important materials, but also in the perspective of the design of new bioinspired or reconstructed biological materials.  As a consequence, an intense experimental, numerical, and theoretical effort has been recently devoted in this field \cite{BY}. From an experimental point of view, a great impulse in this direction has been delivered by new experimental techniques \cite{Rit}, such as Atomic Force Microscopy (AFM) \cite{Rief1}, laser optical tweezers \cite{KSGB}, magnetic tweezers and single molecule fluorescence techniques.
The typical experiments is a mechanically induced unfolding of a macromolecule composed of $n$ unraveling domains, such as a polymeric polypeptide, dextan \cite{R}, silks \cite{Hennady}, proteins (see \cite{Rief1} for the titin), DNA/RNA strands \cite{Smith}.

From a theoretical point of view the thermo-mechanical behavior of multidomains proteins has been undertaken following different approaches: Molecular Dynamics, off lattice models, all atom Montecarlo approaches (see \cite{MS} and references therein), phenomenological approaches, Statistical Mechanics energy landscape analyses,  with funneling \cite{JU} and Inherent Structure models \cite{NP}.
Molecular Dynamics theories \cite{MS} have been restricted by the computational effort required to describe the unfolding of such large macromolecules at the AFM loading time scale.
On the other hand, the statistical approaches for the discrete chain have been essentially based on numerical techniques, whereas analytical results have been obtained only in the thermodynamic limit hypothesis that hides the crucial role of finite size and discreteness of the unfolding phenomenon.

\begin{figure}[!h]\vspace{0 cm}
\begin{center}  \includegraphics[scale=0.21]{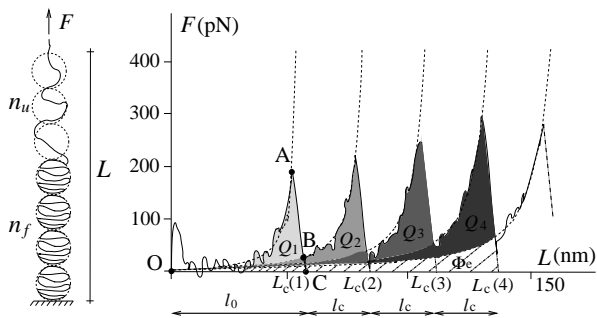} \vspace{-0.2 cm}  \caption{ \label{esp}
Energetic interpretation of a typical unfolding experimental stick-slip AFM force-elongation curve (continuous line) and scheme of the energetic decoupling of the external work into unfolding
(dissipated) energy $Q_i, i=1,2,3,4$, and elastic stored energy $\Phi_e$
(dashed region).
Dashed lines represent approximating WLC curves each characterized by a different contour length $L_c(i)$, $i=1,...,4$; $l_c$ is the (fixed) contour length increase at each unfolding event. }
\end{center}
\end{figure}

To fix the ideas, in Fig.\ref{esp} we schematically show a typical AFM single molecule stretching experiment on an engineering reconstructed macromolecule domain. The figure shows the typical sawtooth force-length diagram, that can be explained as a stick-slip dynamical evolution in a wiggly energy landscape, characterized by multiple energy wells, each corresponding to a given hard-soft microstructure configuration.
Thus, for growing assigned end-to-end length  (see  \cite{Lacks}) the chains alternates `slow' ({\it intrabasin}) steps in energy wells at fixed folded/unfolded configuration, followed by `fast' ({\it interbasins}) transitions corresponding to the unfolding of (tipically {\it single}) crystals. These transitions are signaled by the periodic localized force drops induced by the entropy jumps due to the creation of new free monomers at each  $\beta$-sheet unfolding.

The Molecular Dynamics approach in \cite{Lacks}, based on the Inherent Structure formalism, clarifies that the observed unfolding is regulated by three main timescales: loading time $\tau_{load}$, intrabasin relaxation time $\tau_{intra}$, and  interbasins transition time $\tau_{inter}$. Our theoretical model is based on the time scale separation hypothesis
\begin{equation}\tau_{intra} \ll \tau_{load} \ll \tau_{inter}.\label{ts}\end{equation}
In this time scale regime \cite{Lacks} unfolding results as an alternated sequence of {\it purely elastic}, intrabasin, stick evolutions and {\it purely dissipative}, interbasins, slip transitions, localized at fixed unfolding length thresholds.

In the case of AFM induced unfolding, the {\bf non dissipative hypothesis} of the intrabasin evolution is supported by the observation \cite{DDL, KSGB} that during the (slip)  evolution at fixed folded/unfolded configuration, the behavior is fully reversible, thus indicating that the system relaxes to the local energy minimum.
On the other hand, the { \bf fully dissipative hypothesis} of the `fast' (stick) interbasins transitions,  results by the observations (see Fig.\ref{esp}) that at the AFM loading time scale the  $\beta$-sheet unfolding events are (mainly) localized at fixed macromolecule end-to-end lengths.

We remark that the time scale separation (\ref{ts}) has been successfully adopted in other stick-slip evolutions associated to abrupt microstructure transitions in multyvalley energy landscapes at `low loading rates' and `low temperature regimes' such as depinning or nucleation of new defects, dislocations and Frank--Read sources in metal plasticity (see \cite{Puglisi-Truskinovsky1} and references therein), and Barkhausen jumps in ferromagnetism \cite{Bert}.

Based on energy conservation (Gibbs equation), under our hypothesis (\ref{ts}), the external work (see \cite{Puglisi-Truskinovsky1} for a theoretical discussion) can be  decomposed as follows. By focusing again on the AFM experiment in Fig.\ref{esp}, suppose that we begin stretching the macromolecule from its natural state. The macromolecule follows elastically the first equilibrium path (O--A in Fig.\ref{esp}) with the external work $W$ accumulated as elastic energy $\Phi_e$ ($\Delta W=\Delta \Phi_e$). At the first $\beta$-sheet unfolding (path OA) -- here approximated as instantaneous with no external work $W$ -- there is an internal energy discontinuity $\left [\hspace{-0.5 mm}| \Phi \hspace{-0.0 mm} |\hspace{-0.5 mm}\right ]_1$ (area O--A--B) that by energy conservation equals the unfolding energy $Q_1$ of the first $\beta$ sheet. Similar considerations can be extended to the next elastic and dissipative steps, so that the total dissipation is $Q=\sum Q_i=\sum \left [\hspace{-0.5 mm}| \Phi \hspace{-0.0 mm} |\hspace{-0.5 mm}\right ]_i$.

Based on previous considerations,  here we consider an energetic approach for a  two-states material, inspired by the model in \cite{DeTommasi-Puglisi-Saccomandi, DeTommasi-Puglisi-Saccomandi2}, where the authors describe the hysteresis of filled polymers and spider silks. Similarly, a two-state energetic approach was proposed  in \cite{Buhot-Halperin} to describe the helix$\rightarrow$coil transition of polypeptide chains regulating the damage of multi-block copolymers. In this work the authors obtained an analytic solution in the thermodynamic limit of large number of breakable links. 

In the field of protein mechanics, based on the approach in \cite{Buhot-Halperin}, an important step in the comprehension of the energy competition between the unfolding and entropic energy terms, has been delivered in  \cite{Makarov}. In this paper, the author models the unfolding of a biomacromolecule as a chain composed of folded and unfolded domains, both elastic with Gaussian type response, combined with an Ising-like unfolding energy. The resulting MD simulation well describes the unfolding effect in a protein macromolecule,  whereas analytical results are obtained only in the thermodynamic limit of many folded domains \cite{Makarov}. 

More recently, in \cite{Staple} a Statistical Mechanics based model for the stretching of titin proteins has been proposed, considering also the influence of the AFM loading device. A simple Ising model, neglecting the elasticity of both folded and unfolded domains in protein macromolecules, was instead proposed  in \cite{TK} to describe the statistics of unfolding events.
The described competition between the entropic energy of the unfolded fraction and the $\beta$-sheets unfolding energy has been analyzed in \cite{Rief3} and \cite{RBT} via  Monte Carlo simulations combining a Worm Like Chain (WLC) with a two-state Bell type model for the unfolding.
Finally, we recall the fully phenomenological continuum approaches  recently proposed in \cite{Benichou} and \cite{Raj-Purhoit} where the authors show the possibility of describing the protein unfolding as an energy minimization of a continuum system with a non-convex internal energy. In particular in \cite{Benichou},  based on the general approach of \cite{Puglisi-Truskinovsky1} and \cite{Puglisi-Truskinovsky2} for the description of the mechanical behavior of a bi-stable discrete chains, the authors obtain an interesting characterization of an optimality condition of the number of $\beta$-sheet domains with respect to the toughness of the macromolecule.

Here, based on previous hypotheses, we obtain a fully analytic rate-independent dissipative lattice, describing the stretch-induced unfolding of macromolecules. Interestingly we show the existence of a fundamental experimentally measurable non-dimensional parameter $\xi$ defined in (\ref{xi}), representing the ratio of the elastic and unfolding energy of the single folded domains, that regulates the dissipation and the unfolding thresholds of each folded/unfolded phase configuration. 
The model is also extended to describe the possibility of the existence of a hierarchy of variable unfolding energies of the different folded domains. As we show, the effective distribution of unfolding energies, can be deduced on the base of the previously described energetic analysis, using the experimental force-elongation diagrams. Interestingly, the analysis of the behavior of different macromolecules suggest simple phenomenological linear distributions.

Aimed at the possible deduction of a three-dimensional continuum extension of the proposed model for biological tissues, that will be the subject of our future work, we also deduce the continuum limit of the proposed discrete model. Indeed, following statistical approaches, as proposed in \cite{QOB}, the continuum limit model constitutes the base of a multiscale model for the description of the mechanical behavior of networks of modular macromolecules. 
The main advantage of our theoretical approach is that the behavior of the single chain also in the continuum limit is regulated by the experimentally measurable parameter $\xi$ that fully characterizes the unfolding behavior of the macromolecule.

Finally, as an explicit example, we focus on the AFM experiments of titin unfolding. As we show, under a simplifying assumption of rate-dependent effective unfolding energies and on the base of the phenomenological law of the unfolding energy hierarchy recalled above, the analytical model quantitatively well describes the experimental behavior.

\section{Energetic assumptions}\label{ea}
Because in macromolecules $\beta$-sheets crystals the unfolding is typically an all-or-none transition, as confirmed also from the size of periodicity of the experimental unfolding lengths  \cite{Rief1}, we model the molecule as a discrete lattice of $n$ two-states (rigid-folded/entropic-unfolded) links (see the scheme in Fig.\ref{esp}). The folded/unfolded state of the chain is assigned by a set of internal variables $ \chi_i$, $i=1,...,n$, such that $\chi_i= 0$ ($\chi_i=1$) denotes the  folded (unfolded) state. Thus in particular $n_u=\sum_{i=1}^n \chi_i$ is the number of unfolded elements and
$n_f=n-n_u$ is the number of folded elements.

As in the case of Freely Jointed Chain or Worm Like Chain models (see {\it e.g.} \cite{Buhot-Halperinb}), we characterize the behavior of each unfolded link through
its {\it contour length} $l_c$ and end-to-end length $l$, with a free energy density (energy per unit length) $\varphi_e=\varphi_e(\eta)$, where $\eta:=\frac{l}{l^c}$ represents a strain measure. We assume then the limit extensibility condition  $\varphi_e\rightarrow \infty$ as $\eta \rightarrow 1$.

By neglecting non-local interactions (weak interaction hypothesis), the total elastic energy $\Phi_e=\sum_{i=1}^n \chi_i l^c_i \varphi_e\left (\frac{l_i}{l^c_i}\right )$ can be simply expressed as \begin{equation}\Phi_e=L_c\varphi_e (\eta).\label{TM}\end{equation}
Here,
$$\eta=\eta(L,n_u)=\frac{L}{L_c(n_u)}$$
is the strain in the unfolded domain,
$$L=\sum_i \chi_i l_i$$ is (by neglecting the extension of the unfolded domains) the total end-to-end length, and
 \begin{equation}
 \label{lcc} L_c=L_c(n_u) =L_0+n_u l_c \end{equation}
is the total contour length of the chain. In (\ref{lcc}) $L_0$ denotes the `initial' (virgin) contour length of the unfolded domain and $l_c$ the contour length of each unfolded domain.

Indeed at equilibrium, by neglecting the elastic energy of the (rigid) folded domains, the total elastic energy is given by $\Phi_e=\sum_{i=1}^{n} \chi_i l_c^i \varphi_e(\frac{l_i}{l_c^i})$, where $l_c^i$ and $l_i$ are the (possibly variable)  contour length and end-to-end length of the $i$-th unfolded link, i.e. for all $i=1,...,n$ with $\chi_i=1$. Under an equilibrium hypothesis we have a constant stress for all unfolded links, i.e. $l_c^i \frac{ d \varphi_e(l_i/l_c^i)}{d l_i}=F$. Thus, for a convex energy density (monotonic derivative $d\varphi_e/dl$), such as WLC or FLC, the strain is homogeneous in all unfolded elements, i.e. $\eta_i=\frac{l_i}{l_c^i}=\eta=\frac{L}{L_c}$, for all $i=1,...,n$ with $\chi_i=1$. Thus we have $\Phi^e=\sum l_c^i \varphi^e(\eta)=L_c \varphi^e(\frac{L}{L_c})$.

 Following \cite{Buhot-Halperin} and \cite{Makarov} here we consider an Ising type unfolding energy
$$\begin{array}{lll}\Phi_{tr}&=& -\sum_{i=1}^{n} (Q-J) (1-\chi_{i})- \vspace{0.2 cm}\\ &  & J \sum_{i=1}^{n-1} (1-\chi_{i}) (1-\chi_{i+1})=Q(n-n_u) + J n_{b_f},\end{array}$$ depending on the internal variables $\chi_i$ and the number  $n_{b_f}$ of contiguous folded blocks in the folded/unfolded configuration. Here $Q$ is the unfolding energy for a single domain and $J$ is a penalizing `interfacial' energy term (measuring the loss of internal energy due to the unbind terminal $H$-bonds of each contiguous folded domain \cite{Buhot-Halperin}).

To get the total energy $\Phi_{tot}=-k_BT \ln [p(L, n_u, n_{b_f})]$ (where $T$ is the temperature and $k_B$ is the Boltzmann constant),
we have to know the probability $p(L, n_u, n_{b_f})$ of a given elastic configuration of the chain with a microstructure corresponding to $n_u$ and $n_{b_f}$. In particular, we have $p(L, n_u, n_{b_f})=\Omega(n_u, n_{b_f})p_{e}(L, n_u)p_{tr}(n_u, n_{b_f})$, where $\Omega(n_u, n_{b_f})$ represents the number of sequences with assigned $n_u$ and $n_{b_f}$,
$p_{e}(L, n_u)\sim \exp (-\frac{\Phi_{e}(L, n_u)}{k_B T})$ represents the probability of attaining a length $L$ at given $n_u$ and $p_{tr}(n_u, n_{b_f})\sim \exp (-\frac{\Phi_{tr}(n_u)}{k_B T})$ is the probability of a state with assigned $n_u$.
So, we obtain $\Phi_{tot} =\Phi_e(L, n_u)+\Phi_{tr}(n_u) -  T S(n_u, n_{b_f})$
where $S(n_u, n_{b_f})=k_B \ln \Omega(n_u,n_{b_f})$ represents the mixing entropy term.

Observe that the coupling energy term $J$ penalizes the multiplicity of folded blocks, whereas the mixing entropy term induces multi domains configurations. In the following we assume, as in \cite{Buhot-Halperin},  that the penalizing term $J$ dominates this effect, so that we always consider single folded domains configurations ({\it i.e.} we assume $n_{b_f}=1$, known as di-block approximation). This hypothesis is supported by the MD simulations \cite{HSLS} showing an unfolding strategy with always one single connected internal unfolded domain inside two boundary folded domains.

Under these hypotheses we obtain the simple expression of the total energy
\begin{equation} \Phi_{tot}= \Phi_e(L, n_u)+n_u Q+ \mbox{const}.
\label{fitr}\end{equation}

We remark that to avoid the introduced di-block approximation, not always experimentally verified, one needs to evaluate the partition function ({\it e.g.} \cite{Makarov}, \cite{Staple}) and only numerical results in the discrete model can be obtained. Moreover stochastic processes considering fluctuations in both the unfolding forces \cite{CV} and the unfolding lengths, are possible extensions of the proposed model. Also these extensions require the employment of numerical approaches.

\section{Energy minimization}

Consider a WLC force-length relation proposed in \cite{Marko-Siggia}\begin{equation}
\frac{F }{k_B T}= \frac{1}{4 L_p}\left (\frac{2\eta -\eta^2}{(1-\eta)^2}+\eta \right)
\label{WLC}
\end{equation}
corresponding to an energy density
\begin{equation} \label{wlc}
\varphi_e(\eta)=\frac{k_BT}{4L_{p}} \left( \frac{\eta ^2}{1-\eta } + 2\eta ^2 \right).
\end{equation}
 Thus, using (\ref{TM}), the adimensionalized total entropic energy of the unfolded fraction can be written as
\begin{equation}
 \bar \Phi_e:=\frac {\Phi_e}{k_B T} =
 \frac{1}{4 L_p} \left( \frac{\eta ^2}{1-\eta } + 2\eta ^2 \right) L_c(n_u).
\label{WLC-Energia-n}
\end{equation}

\begin{figure}
\centering
\includegraphics[scale=0.35]{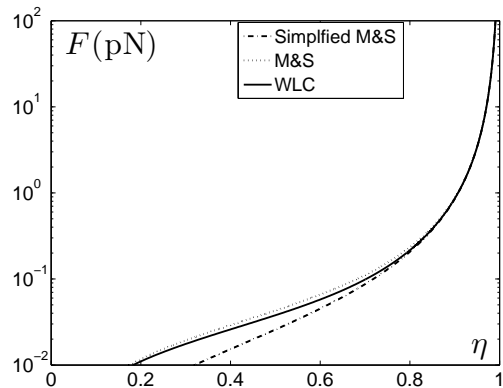}
\caption{Force-strain curve (log scale stresses the differences in the low force regime) for the WLC compared with the usual Marko and Siggia approximation in \cite{Marko-Siggia} and with the simplified model in (\ref{WLCS}) for $L_p=0.42$. Observe that this approximation keeps the same asymptotic behavior as $\l\rightarrow l_c$ of the WLC model in (\ref{WLC-Energia-n}). For low forces ($F<10^-1$ pN) the introduced approximation is significant as compared with the approximation in \cite{Marko-Siggia}; for larger forces the approximation is of the same order of \cite{Marko-Siggia}. \label{MS}}
\end{figure}

In order to attain analytical solutions, we consider the following simplified expression of the WLC energy density: 
\begin{equation} \label{app}
\varphi_e(\eta)=\frac{k_BT}{4L_{p}} \left( \frac{\eta ^2}{1-\eta }  \right).\end{equation}
Observe that this approximation keeps the same asymptotic behavior as $\l\rightarrow l_c$ of the WLC model in (\ref{WLC-Energia-n}). Fig.\ref{MS} shows (in a log scale, stressing the differences at low values of the force) that, while for low forces ($F<10^-1$ pN) the introduced approximation is significant (as compared with the approximation in \cite{Marko-Siggia}), for larger forces the approximation is of the same order of \cite{Marko-Siggia}. Since in the low force regime the elasticity is mainly regulated by the PEVK and tertiary structure elasticity (see \cite{ZE} and \cite{HSLS} for details), this approximation appears inessential in both the qualitative and the quantitative analysis of the behavior during the large-forces unfolding regime of interest for titin unfolding. Moreover, we remark that the approximation (\ref{WLC}) has been shown to be inefficient in the low force regime in \cite{DC}, where the authors introduce a Mooney Rivlin type correction to the WLC constitutive law.

Thus, using (\ref{TM}) and (\ref{lcc}) the total elastic energy is
$$\Phi_e(\eta,n_u)=\frac{ k_B T }{4L_{p}}  \frac{\eta ^2}{1-\eta }L_c(n_u) $$
and, correspondingly, the total force-deformation relation is
\begin{equation}
F(\eta,n_u)= \frac{ k_B T }{4L_{p}}\frac{2\eta -\eta^2}{(1-\eta)^2}.
\label{WLCS}
\end{equation}
Finally, according with (\ref{fitr}), the total  energy is
\begin{equation}\label{ENERS}\Phi_{tot}=\frac{ k_B T }{4L_{p}}  \frac{\eta ^2}{1-\eta }L_c(n_u) + Q n_u.\end{equation}
\begin{figure}[!h]\hspace{-0.1 cm}
\includegraphics[scale=0.42]{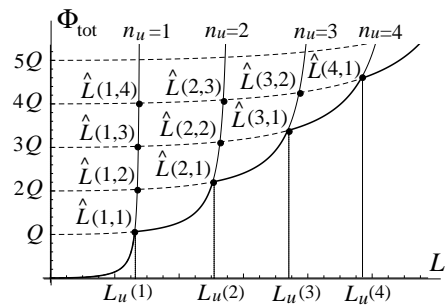}\vspace{-0.1  cm}
\caption{ Scheme of the energy minimization: with bold line we represent stable (global energy minimum) solutions. } \label{ttt}
\end{figure}

We follow a Griffith-like approach \cite{Gri}, minimizing the total unfolding (fracture) energy plus elastic (entropic) energy, and based on (\ref{ts}) we assume that the observed solutions are the global minima of $\Phi_{tot}$ in (\ref{ENERS}).

The first important step is to justify the experimental observation that the $\beta$-sheets unfold one at a time, resulting in a constant increase of the contour length  (see \cite{Rief1}). To obtain this result, we begin by evaluating the solution of the equations  $\Phi_{tot}(L,n_u+m)-\Phi_{tot}(L,n_u)=0$ and get the intersection lengths $L=\hat L(n_u, m)$   (see Fig.\ref{ttt}).
The searched result  follows by the observation that $ \frac{\partial\hat L(n_u,m)}{\partial m}>0$
so that if $\bar n_u$ is the branch corresponding to the global minimum, by increasing $L$ it looses its
global stability at the intersection with the equilibrium branch $\bar n_u+1$ (see Fig.\ref{ttt}).  Thus the chain unfolds with a sequence of single $\beta$-sheets unfolding at the threshold assigned by $
\Phi_e(\eta, n_u) - \Phi_e(\eta, n_u+1)= Q, \hspace{0.6 cm} n_u=0,...,n-1,
$:
\begin{equation} L_u(n_u)= \hat L(n_u,1)=\frac{2 L_c+l_c
-  \!\sqrt{2 L_c(L_c+l_c)/ \xi+ l_c^2}}{ (2-1/\xi)}.
\label{lthr}\end{equation}
 (in this formula and in the following we omit the $n_u$ dependence of $L_c$).
 
 In
 (\ref{lthr}) we introduced the main   non-dimensional parameter of the model  $\xi$ \begin{equation} \xi=\frac{8 L_p   }{ k_B T}\frac{Q}{l_c}	\label{xi} \end{equation}
representing a measure of the ratio between the elastic and fracture energy of the single $\beta$-sheet. Indeed we observe that according with (\ref{wlc}) we have that $\frac{ k_B T l_c}{8 L_p   }= \varphi_e(\frac{1}{2}) l_c$ measuring the elastic energy of a single domain when the deformation is a half of the maximum elongation (contour length).

It is easy to verify that $L_u\in (0, L_c)$ and that
$\frac{d L_u}{dn_u}>0$.  As a result, the $n_u$ branch corresponds to the global energy minimum for
$$ L \in (L_u(n_u-1),L_u(n_u)), n_u\in(1,n-1),$$
representing the existence domain of the $n_u$ branch under our energy minimization hypothesis.
In the special cases of the virgin curve, with $n_u=0$, we have $L \in(0,L_u(0))$ and of the fully unfolded chain, with  $n_u=n$, we have $L \in(L_u(n-1),L_r)$, where $L_r$ is the fracture threshold of the fully unfolded chain.

Using (\ref{WLCS}), we get the unfolding force $F_u=F(L_u/L_c,n_u)$
\begin{equation}
F_u\!\!=\!\!\frac{k_B T}{ 4 L_p}\!\!\left(\!\! \left (\!\!\frac{2 (\xi-1)L_c}{l_c\, \xi+2L_c-\sqrt{(l_c\,\xi+2L_c)^2+4(\xi-1)L_c^2}} \!\! \right)^2\!\!\!\!-1 \!\!\right).\label{fthr}
\end{equation}

Observe that using (\ref{lthr}) and (\ref{fthr}) it is also possible to obtain an explicit relation between the unfolding forces and the unfolding end-to-end lengths
\begin{equation}
F_u(L_u)=\frac{k_B T}{  L_p}\frac{ L_u\left( L_u+\sqrt{8 L_u^2/\xi +l_c^2}-l_c \right)}{\left ( \sqrt{8L_u^2/\xi+l_c^2}-l_c \right )^2}.
\label{continuo-analitico}
\end{equation}
It is important to observe that since $\frac{d  F_u}{d n_u}<0,$
under the hypothesis of fixed unfolding energy $Q$ of the different $\beta$ sheets, the system shows a softening behavior during the unfolding.

\begin{figure}[!h]
\begin{center}
\includegraphics[scale=0.25]{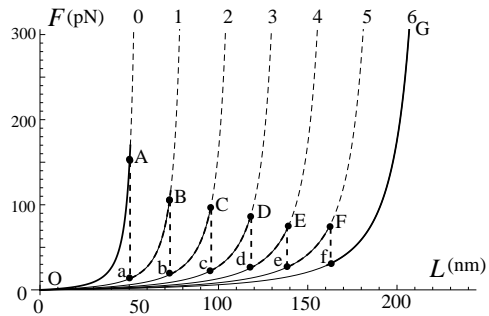}
\caption{Unfolding behavior for a system of $n=6$ initial folded domains. Here  we considered the parameters: $l_o=58$ nm, $l_{c}=28.43$ nm, $l_{p}=0.36$ nm, $Q=770$,  $\Delta Q=420$. Each equilibrium path is labelled by the number $n_u$ of unfolded domains. \label{modelsoft}
}
\end{center}
\end{figure}

The stretch induced unfolding of the system is shown with bold line in Fig.\ref{modelsoft}. As described before, the system reproduces the typical experimental behavior of unravelling macromolecules with a regularly spaced sequence of unfolding events of the hard domains. Thus if we start loading from the virgin configuration ($n_u=0$, point O in the figure), the system follows elastically the equilibrium curve $n_u=0$, until the unfolding energy $Q$ equals the jump of the entropic energy due to the transition from the branch $n_u=0$ to the branch $n_u=1$ (path A-a in the figure). By increasing further the assigned length the system follows the new branch until another sudden transition to the branch $n_u=2$ is observed when it becomes energetically favorable (path B-b). Similar transitions with single domains unfoldings are then observed, until all the crystals unfold and the system shows a hardening behavior due to the entropic elasticity of the fully unfolded chain (curve f-G in the figure).

In particular we observe that the theoretical model shows a softening behavior during the unfolding regime, with the unfolding force decreasing with the number $n_u$ of unfolded domains.
 We remark that the experimental behavior of stretch induced unfolding of macromolecules show a variable behavior, with typically nearly constant thresholds (see {e.g.} \cite{Fisher}, but with increasing, decreasing or random transition thresholds (see {e.g.} \cite{Rief1} for titin macromolecule and \cite{Lv} for artificial elastomeric protein). 
 
 In the following section we discuss this issue and propose an extension of the proposed model able to reproduce the hardening effect observed {\it e.g.} in titin macromolecule unfolding.

\section{Unfolding Energy Hierarchy}

The observed variability in the unfolding experiments (force plateaux, hardening, softening) can be addressed to a hierarchy of unfolding energies of the crystals, due to inhomogeneity effects with variable properties of the crystal domains \cite{BP, Rief1, Rief3}. Indeed the experiments show an inhomogeneity of unfolding with a variable bond-breaking barriers \cite{Staple} possibly due to interfacial energy effects \cite{Titin-1}. Moreover another important effect can be due to the different orientation of the crystals in the macromolecule. Indeed the crystals show a directional deformation
response of the folded domains, that, according with the loading direction follow different paths in the energy landscape, leading to different unfolding forces and energies \cite{Dietz}.
Another effect, inducing hardening, is the so called $n$-effect (see \cite{Dietz})  that, based on statistical considerations, addresses the observed hardening to a progressively reduced number of folded crystals available for unfolding in the macromolecule for growing elongations. 

\begin{figure}[!h]\vspace{0.5 cm}\hspace{-0.4 cm}
\includegraphics[scale=0.295]{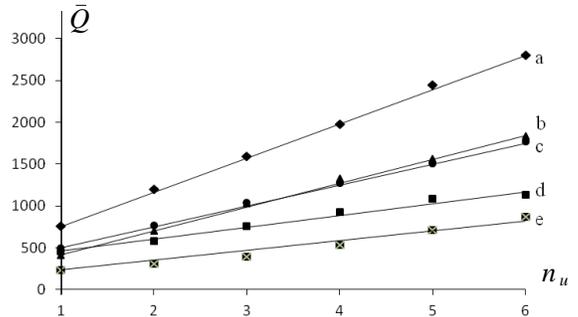}\vspace{-1 cm}
\caption{ \label{esperimenti1}
Unfolding energies as a function of $n_u$ deduced from the following experiments: a) AFM experiment on Titin from \cite{Rief1}; b) AFM experiment on TNfnAll protein from \cite{Oberhauser}; c) AFM experiment on Titin from \cite{CV}; d) AFM experiment on Tenascin-C from \cite{Fisher}; e) AFM experiment on Titin from \cite{Linke}. }
\end{figure}

To take care of these experimental effect in the following we consider the possibility of variable unfolding energy of the hard domains.  Thus,  we first observe that, following the analysis of previous sections, based on the experimental force-displacement unfolding diagrams, we may estimate the fracture energy of each unfolding event using the relation
\begin{equation}\Phi_e(\eta_u, n_u) - \Phi_e(\eta_u, n_u+1)= \bar Q (n_u)\label{veb} \end{equation} where $\bar Q(n_u)$ represents the variable fracture energy of the $n_u$-th $\beta$-sheet and $\eta_u=L_u(n_u)/L_c(n_u)$ is the strain corresponding to the unfolding threshold of the $n_u$ configuration. Based on this relation  we analyzed the experimental length-force diagram for different unfolding macromolecules: for titin in \cite{Rief1}, \cite{CV}, and \cite{Linke}, and for TNfnAll protein from \cite{Oberhauser} and Tenascin-C from \cite{Fisher}. The results are described in Fig.\ref{esperimenti1} and interestingly show a linear empirical law
\begin{equation}
 \bar Q({n_u}) =  Q + (n_u-1) \triangle  Q,
\label{WLC-barriera1}
\end{equation}
where $Q=\bar Q(1)$ represents the fracture energy of the weakest folded domain, that has the important role of regulating the stability of the initial unfolded configuration and the initial unfolding length $L_u(1)$, whereas $\Delta Q$ is a fixed energy increment for successive unfolding events (see Fig. \ref{esperimenti1}).

It is important to remark that in the case of increasing unfolding energies, the di-block approximation can fail, with the order of unfolding that is no more regulated by the interfacial energy effects discussed in Sect.\ref{ea}. In this case an exact solution of the problem requires a numerical analysis. 
To keep an analytical treatment also in the case of variable unfolding energies, we here suppose that both the mixing entropy contribution and the interfacial energy term are negligible as compared with the unfolding energy increment $\Delta Q$. Under this simplifying assumption, we may first easily extend the considerations described in Fig. \ref {ttt} to obtain that the domains unfold one at a time in the order of their unfolding energies. Then we may again explicitly evaluate the unfolding lengths (\ref{lthr}) and forces (\ref{fthr}) by simply using (\ref{lthr}) with a variable parameter
\begin{equation}\label{xixi}\xi(n_u)=  \frac{8 L_p   }{k_B T l_c}Q(n_u)\end{equation}
measuring the variable ratio of dissipated and elastic energy of the $\beta$-sheets.

\section{An explicit example: Titin unfolding}

To show the feasibility of the proposed model in quantitatively predicting the experimental behavior of macromolecule unfolding, in this section we analyze the diffusely studied AFM stretching experiments of titin, the protein responsible of the passive strength of muscles. These proteins are very long macromolecules with contour length larger than 1 $\mu$m \cite{Wang}, whose secondary structure is characterized by the presence of immunoglobulin (Ig) and fibronectin type III (FNIII) domains, folded in forms of $\beta$-sheets, connected to the PEVK domain (rich in proline, glutamate, valine and lysine, see {\it e.g.} \cite{LHMS}). At low forces the elasticity is regulated by the tertiary structure and the  (random coil) domain orientation, combined with the elasticity of PEVK domains \cite{LHMS, Titin-1, HSLS}. At higher forces the macromolecule response is dominated by an energetic competition of the entropic elasticity of the unfolded fraction of (Ig) and (FNIII) domains and by the hentalpic contribution of the folded$\rightarrow$unfolded transition of $\beta$-sheets.

\begin{figure}[!h]\vspace{-0.3 cm}
\includegraphics[scale=0.32]{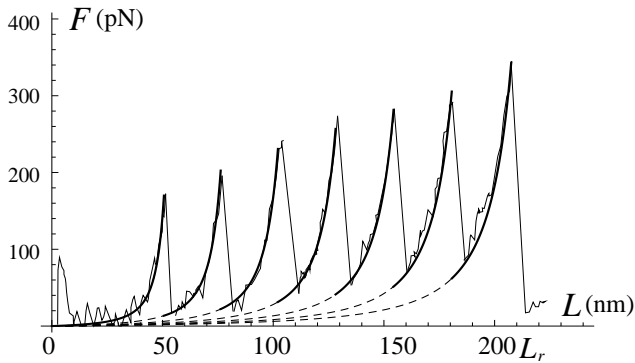}\vspace{-0.4 cm} \\
\caption{
Comparison between the AFM experiment for the titin protein reproduced from \cite{Rief1} (continuous curves) with the Force vs. Elongation curves deduced by  (\ref{WLCS}), (\ref{lthr}), (\ref{xixi}), and (\ref{WLC-barriera1}). Here we considered the parameters: $l_o=58$ nm, $l_{c}=28.43$ nm, $l_{p}=0.36$ nm, $Q=770$,  $\Delta Q=420$.\label{exptheor}}
\end{figure}

As anticipated in the Introduction, one important effect in the case of titin is the rate dependent behavior of the unfolding events (see {\it e.g.} \cite{Rief1}) requiring a MD approach as described  in \cite{Lacks} or a reaction theory approach \cite{ER}. Here, with the aim of deducing a fully quantitative representative analytical approach, following \cite{Lacks, Bry, BY, RBT}, we take care of  the observed rate-dependence (i.e. rate dependent unfolding energy barriers and dissipation), by considering {\it effective}, rate-dependent dissipation energies $Q_i$. Moreover, since typically  no refolding is detected during unloading  \cite{Rief5}, we assume that the (hard-soft) transitions as irreversible.

In Fig.\ref{exptheor} we show the ability of the model in describing quantitavely the behavior of titin unfolding experiments reported in \cite{Rief1}, based on the 
relations (\ref{WLCS}), (\ref{lthr}), and (\ref{xixi}), by using the empirical law  (\ref{WLC-barriera1})  for the variable unfolding energy with the values deduced by \cite{Rief1} and reported in Fig.\ref{esperimenti1}$_a$.

\section{Continuum Limit}

In this section, aimed to a deduction of a continuum model for protein materials (see {\it e.g.} \cite{DeTommasi-Puglisi-Saccomandi2}), we analyze the continuum limit of the proposed model, obtained as a limit when $n\rightarrow \infty$. We consider this limit at fixed total unfolded length that using (\ref{lcc}) is given by
$$\bar  L_c=L_0+n l_c,$$
({\it i.e.} $l_c$ decreases as $n$ grows). To this end we introduce the unfolded fraction 
$$\nu_u:=\frac{n_u}{n},$$
considered here as a (damage) continuum internal variable, with $\nu_u\in (0,1)$ and 
$\nu_u=0$ in the virgin state and $\nu_u=1$ in the fully unfolded state. The total contour length is then in the continuum case a function of the unfolding fraction that, using (\ref{lcc}), is given by
$$L_c=L_c(\nu_u) =L_0+\nu_u (\bar L_c-L_0).$$
Thus the damage variable $\nu$ measures the change of contour length with in particular $L_c=L_0$ in the virgin configuration ($\nu_u=0$) and $L_c=\bar L_c$ in the damage saturation of the fully unfolded state ($\nu_u=1$).

The total energy, using (\ref{ENERS}), can be rephrased as 
\begin{equation}\Phi_{tot}=\hat \Phi_{tot}(\nu)=\frac{ k_B T }{4L_{p}}  \frac{\eta ^2}{1-\eta }L_c(\nu_u) + \bar Q \nu_u, \label{ENERSCONT}\end{equation}
where we introduced the rescaled  expression
$$\bar Q:=n Q$$
ensuring that the unfolding energy $\bar Q$ decreases with growing $n$ and so with decreasing $l_c$. Here the deformation variable depends on the continuum damage variable $\nu_u$ according with the following relation 
\begin{equation}\eta=\eta(L,\nu_u)=\frac{L}{L_c(\nu_u)}.\label{etac}\end{equation}

\begin{figure}
\begin{center}\vspace{0.4 cm}
\includegraphics[scale=0.3]{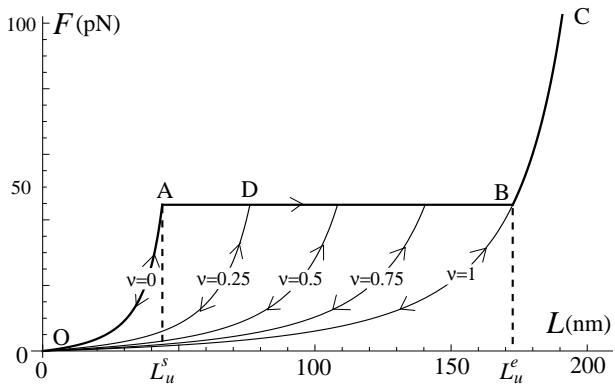}
\caption{Unfolding behavior of the continuum limit model. Bold lines represent the loading path, whereas continuous lines the unloading paths at different values of the unfolded fraction $\nu_u$. Here we used the same parameters of Fig. \ref{modelsoft}.
\label{contin}
}
\end{center}
\end{figure}

The obtained framework can be inscribed in the classical variational approach for damage known as pseudoelasticity \cite{DO}, requiring the minimization of a damage dependent energy. We refer the reader to \cite{DeTommasi-Puglisi-Saccomandi} for a detailed discussion of this approach. Based on our irreversibility assumption (no refolding), if we indicate by $L_{max}$ the maximum attained assigned length we have the following behavior. To determine the global minimum of the energy (\ref{ENERSCONT}), during loading ($L=L_{max}$) we minimize both with respect to $L$ and $\nu_u$. Minimization with respect to $L$ (that is $\frac{\partial \hat \Phi_{tot}(L,\nu_u)}{\partial L}=0$) delivers the equilibrium force as in (\ref{WLCS}) with the deformation variable defined in (\ref{etac}). The minimization with respect to $\nu$ (that is $\frac{\partial \hat \Phi_{tot}(L,\nu_u)}{\partial \nu_u}=0$) delivers the damage as a function of the assigned length
\begin{equation}\nu_u=\bar \nu(L)=\frac{\left(1+\sqrt{\frac{2}{\xi }}\right)L-L_0}{\bar Lc-L_0}.\label{nuc}\end{equation}
Interestingly in this limit we obtain a constant unfolding force (plateau)
$$F_u=\frac{K_bT}{ L_p} \left(\frac{\xi }{8}+ \sqrt{ \frac{\xi}{8}}\right).$$
Using (\ref{nuc}) we obtain that the unfolding begins ($\nu_u=0$) at 
$$L=L_u^s= \frac{\sqrt{\xi}L_0}{\left(\sqrt{\xi}+\sqrt{2}\right)}$$
whereas the fully unfolded state ($\nu_u=1$) is attained at 
$$L=L_u^e=\frac{\sqrt{\xi}\bar L_c}{\left(\sqrt{\xi}+\sqrt{2}\right) }.$$
The obtained unfolding behavior is shown in Fig.\ref{contin} (path OABC).

During unloading ($L<L_{max}$), since we neglect refolding, the behavior is again given by (\ref{WLCS}) with fixed damage that by (\ref{nuc}) is given by $\nu_u=\bar \nu_u(L_{max}).$ Different unolading paths are shown in Fig.\ref{contin} ({\it e.g.} path DO is attained for an unloading at $\nu_u=0.25$).

\section{Conclusions}

In this paper we propose, based on the time scale separation (\ref{ts}), an energetic model for the description of the important phenomenon of stretch induced unfolding of macromolecules. By considering a di-block approximation and by neglecting rate-dependent effects (possibly considering rate-dependent effective unfolding energies) we deduced a fully analytical model delivering the unfolding forces and lengths for the different equilibrium branches, all depending on the deduced main dimensional parameter $\xi$. 
The results have also been extended to the case of variable unfolding energies,
based on the empirical law (\ref{WLC-barriera1}) that we deduced by the experiments. Despite the adopted simplifying hypothesis, the deduced analytical model shows a good qualitative (stick-slip unfolding evolution with regular spacing of the localized unfolding events) and quantitative agreement (see Fig.\ref{exptheor}) with the experimental behavior. 
We also showed that our approach is amenable of the thermodynamic, continuum, limit useful for the extension
of the model to the analysis of network of modular macromolecules \cite{QOB}
that will be the subject of our future studies.


\begin{thebibliography}{00}

\bibitem{Ack}
AT. Ackbarov et al, Proceedings of the National Academy
of Science, 104, 42, 16410-16415 (2007).

\bibitem{Benichou} I. Benichou, S. Givli, Appl. Phys. Lett., 98, 091904 (2011).

\bibitem{Bert} G. Bertotti  \emph{Hysteresis in magnetism}, Academic Press, Boston (1998).


\bibitem{Borisov1}
O. V. Borisov et al, Europhysic Letters, 34, 9, 657-662 (1996).

\bibitem{Borisov2}
O. V. Borisov et al, Macromol. Symp., 113, 11-17 (1997).


\bibitem   {BP} D. J. Brochwell et al., Nat. Struct. Biol., 10, 731-737 (2003).

\bibitem{Bry} J.D. Bryngelson et al., Proteins, 21, 167--195 (1995)


\bibitem{BY} M.J. Buehler, Y.C. Yung, Nat. Mater., 8, 175--188 (2009).

\bibitem {Buhot-Halperin} A. Buhot, A. Halperin, Physical Review Letters, 84, 2160  (2000).

\bibitem {Buhot-Halperinb} A. Buhot, A. Halperin, Macromol., 35, 3238-52 (2002).


\bibitem {DeTommasi-Puglisi-Saccomandi} D. De Tommasi, G. Puglisi,
G. Saccomandi, Physical Review Letters, 100, 085502 (2008).

\bibitem{DeTommasi-Puglisi-Saccomandi2} D. De Tommasi, G. Puglisi,
G. Saccomandi, Biophys. J., 98, 1941-1948 (2010).


\bibitem {DeTommasi-Puglisi-Saccomandi3} D. De Tommasi, G. Puglisi,
G. Saccomandi, J. Rheology, 504, 495 (2006).

\bibitem{Dietz}
Dietz et al, Proc. Nat.
Acad. Sci. 103, 12724-12728 (2006).

\bibitem{DC}
A.V. Dobrynin and J.Y. Carrillo, Macromol., 44, 140--146 (2011)

\bibitem{DO}
A. Dorfmann, R. W. Ogden, Int. J. Sol. Struct., 40, 2699--2714 (2003).

\bibitem{DDL} N. Duff, N.H. Duong, D.J. Lacks, Biophys. J., 91, 3446--3455 (2006).

\bibitem{ER} E. Evans, K. Ritchie, Biophys. J., 72, 1541--1555 (1997)

\bibitem{ERb} E. Evans, K. Ritchie, Biophys. J., 76, 2439--2447 (1999)

\bibitem   {Fisher} T. E. Fisher, A. F. Oberhauser, M. Carrion-Vazquez, P. E. Marszalek, J. M. Fernandez, Trend Biochem. Sci., 24, 379-384 (1999).

\bibitem{Gri}
A. A. Griffith, Phil. Trans. R. Soc. A 221, 163 (1920).

\bibitem{Harrington}
M. J. Harrington et al, J. R. Soc. Interface, 2913 (2012).

\bibitem{HSLS}  J. Hsin, J. StrŸmpfer, E.H. Lee, K. Schulten, Ann. Rev. Biophys.,40, 187-203 (2011).


\bibitem{JU} S.K. Jha, J.B. Udgaonkar, Curr. Science, 99, 457--475 (2010).

\bibitem{TK} T. Kajander, A.L. Cortajarena , E.R.G. Main, S.G. J. Mochrie , and L. Regan, J. Am. Chem. Soc.,  127, 10188--10190 (2005).

\bibitem{KM}
Kazakevicuite-Makovska et al, Proc. Eng. 10, 2597 (2011).

\bibitem{KSGB} M.S.Z. Kellermayer, S.B. Smith, H.L. Granzier, C. Bustamante,
Science 276, 1112 (1997).

\bibitem{Lab}
S. Labeit et al, Science, 270, 293, (1995).

\bibitem{Lacks} D. J. Lacks, Biophys. J., 88, 3494-3501 (2005).

\bibitem{LHMS} E.H. Lee, J. Hsin, O. Mayans, K. Shulten, Biophys. J., 93, 1719--1737 (2007).

\bibitem{LG}W.A. Linke, A. Gr$\ddot{\rm u}$tzner, Pfluger Arch. - Eur. J. Physiol. 456, 101--115 (2008).


\bibitem{Lv}
S. Lv et al., Nature, 465, (2006).

\bibitem {Makarov} D. E. Makarov, Biophys. J., 96, 2160-2167 (2009).

\bibitem{Manca2}
Manca et al, J. Chem. Phys., 136,154906 (2012).

\bibitem{CV}Mariano Carrion-Vazquez, et al., PNAS, 96, 3694Ð3699 (1999).


\bibitem{MMHT} J. L. Mar\'in , J. Mun\~iz, M. Huerta, X. Trujillo,
Gen. Physiol. Biophys., 18, 30--309 (1999).

\bibitem{Marko-Siggia} J.F. Marko, E.D. Siggia, Macromol., 28, 8759--8770 (1995).

\bibitem{MS} L. Mirny, E. Shakhnovich, Annu. Rev. Biophys. Struct., 30, 361--396 (2001).

\bibitem{Miserez}
A. Miserez et al, Nat. Mater. 8, 910 (2009).


\bibitem{NP} N. Nakagawa, M. Peyrard, PNAS, 103, 5279Ð5284 (2006).

\bibitem   {Oberhauser} A. F. Oberhauser, P. R. Marszalek, H. P. Erickson, J. M. Fernandez, Nature, 393, 181 (1998).

\bibitem   {Oberhauser1} A. F. Oberhauser, P. K. Hansma, M. Carrion-Vazquez, J. M. Fernandez,  PNAS, 98, 468-472 (2001).

\bibitem   {Titin-6} F. Oesterhelt, D. Oesterhelt, M. Pfeiffer,  A. Engel, H. E. Gaub, D. J. Muller, Science, 288, 143-146  (2000).



\bibitem{Oroudjev}
E. Oroudjev et al, PNAS 99, 6460 (2002).

\bibitem   {Puglisi-Truskinovsky2}  G. Puglisi, L. Truskinovsky, Cont. Mech. Therm., 14, 437--457 (2002).


\bibitem{Puglisi-Truskinovsky1} G. Puglisi, L. Truskinovsky, J. Mech Phys. Soli., 53 , 655--679 (2005).

\bibitem{Bueler}
Z. Qin and M. J. Buehler, Physical Review E., 82, 061906 (2010).

\bibitem{QOB} 
H.J. Qui, C. Ortiz, M.C. Boyce, J. Eng. Mat. Tech., 128, 509--518 (2006).

\bibitem{Raj-Purhoit}R. Raj, P.K. Purhoit, J. Mech. Phys. Sol., 59, 10054-69 (2011)

\bibitem{R}
M. Rief et al, Science, 275, 28 (1997).

\bibitem{Rief3} M. Rief, J. M. Fernandez, H. E. Gaub,  Physical Review Letters, 81, 4764 (1998).

\bibitem{Rief1} M. Rief, M. Gautel, F. Oesterhelt, J. M. Fernandez, H. E. Gaub, Science, 276 (1997).


\bibitem{Rief5} M. Rief, H. Grubm$\ddot{\rm u}$ller, Chemphyschem, 3, 255-261 (2002).



\bibitem{Rit} F. Ritort. J. Phys, Cond. Matt., 18, R531-R583 (2006).

\bibitem   {RBT} F. Ritort, C. Bustamante, I. Tinoco,  PNAS, 99, 13544 (2002).

\bibitem{Hennady}
H. Shulha et al, Polymer 47, 5821 (2006).

\bibitem{Smith}
S. M. Smith et al, Science. 271, 795 (1996).

\bibitem{Staple} D. B. Staple, S.H. Payne, A. L. C. Reddin, H. J. Kreuzer, Phys. Rev. Lett.,  101, 248301 (2008).

\bibitem{Termonia}
Y. Termonia, Macromolecules, 27, 7378 (1994).

\bibitem{Wang}
K. Wang. Advan. Biophys. J., 33, 123-134 (1996).

\bibitem{Titin-1} H. Li, Wolfgang A. Linke, Andres F. Oberhauser, M. Carrion-Vazquez, J. G. Kerkvliet, Hui Lu, P. E. Marszalek, Julio M. Fernandez, Nature, 418,  998 (2002).


\bibitem{Linke}Wolfgang A. Linke, et al, J. Struct. Biol., 137, 194--205 (2002).

\bibitem{ZE} B. Zhang, J.S. Evans, Biophys. J., 80, 597--605 (2001).























\end{thebibliography}
\end{document}